\documentclass[12pt]{article}
\usepackage{psfig,epsf}
\usepackage{a4wide,graphicx}
\usepackage{cite}
\def\tstrut{\vrule height2.5ex depth0pt width0pt} 

 \def\tstrut{\vrule height2.5ex depth0pt width0pt} 

\begin{document}

\title{Chiral restoration from pionic atoms?}

\author{
C. Garc\'{\i}a-Recio, J. Nieves \\ 
{\small Departamento de F\'{\i}sica Moderna, Universidad de Granada,
  E-18071 Granada, Spain} \\ 
~\\ 
E. Oset \\ 
{\small Departamento de F\'{\i}sica Te\'orica and IFIC,
Centro Mixto Universidad de Valencia-CSIC,} \\ 
{\small Institutos de
Investigaci\'on de Paterna, Aptd. 22085, 46071 Valencia, Spain}\\ 
}

\date{\today}

\maketitle
\begin{abstract}
  
  We evaluate widths and shifts of pionic atoms using a theoretical
  microscopical potential in which the pion decay constant $f_\pi$ is
  changed by an in--medium density dependent one ( $f_\pi(\rho)$),
  predicted by different partial Chiral restoration calculations.  We
  show that the results obtained for shifts and widths are worse than
  if this mo\-dification were not implemented.  On the other hand, we
  argue that in microscopic many body approaches for the pion
  selfenergy, based on  effective Lagrangians, the mechanisms
  responsible for the change of $f_\pi$ in the medium should be
  auto\-matically incorporated. Therefore, the replacement of  $f_\pi$ by
  $f_\pi(\rho)$ in the many body derivation of the microscopic
  potential would be inappropriate.

\end{abstract}

\section{Introduction}

The value of the quark condensate $<\bar{q}q>$ plays an important role
in chiral dynamics \cite{gasser}. In the presence of a nuclear medium
the value of the condensate drops as a function of the density and the
linear terms in the nuclear density $\rho$ have been derived with
different formalisms. In Refs.~\cite{cohen,drukarev} the Hellmann--Feynman
theorem is used, a mean field approach is used in
Refs.~\cite{brown,thornsson} and the Nambu--Jona--Lasinio model is used in
Refs.~\cite{bernard,hatsuda,reinhardt,huang}. The different formalisms lead
to identical results in the terms linear in the
baryon density, $\rho$, giving  at zero temperature
\begin{equation}
\frac{<\bar{q}q>_\rho}{<\bar{q}q>_{\rho=0}} = 1 -
\frac{\sigma_N}{m_\pi^2 f_\pi^2}\rho+ \cdots \label{eq:qqrho}
\end{equation}
where $\sigma_N$ is the pion-nucleon sigma term.
 
Assuming that the Gellmann--Oakes--Renner (GOR) relationship holds, at
zero temperature, for finite baryon
density~\cite{thornsson},\cite{chanfray},\cite{jose}, one gets for the in-medium 
pion decay constant, $f_\pi (\rho)$, defined from  the time
component of the axial current\footnote{Note that there exists a breaking of
covariance because of the nuclear medium which leads to different
renormalization of the space and time components of the axial vector
current.},
\begin{equation}
f^2_\pi (\rho) = -\frac{m_q}{{m^*_\pi}^2}<\bar{q}q>_\rho+ \cdots\label{eq:frho}
\end{equation} 
to leading order in the average quark mass $m_q = \frac12 (m_u +
m_d)$, where $<\bar{q}q>_\rho$ now stands for the $\rho-$dependent
condensate $<\bar u u + \bar d d>_\rho$. Besides $m^*_\pi$ stands for
the pion mass in the medium. Thus, the dropping of the condensate for
finite densities is interpreted as a dropping of the pion decay
constant $f_\pi(\rho)$, leading to the phenomenon of
partial chiral restoration \cite{hatsuda} in the nuclear medium.

    This dropping of the pion decay constant was used in
 Ref.~\cite{wolfram} to suggest that it could solve
 the long standing puzzle of the missing repulsion in the $s-$wave
 pion selfenergy: it might account for the discrepancy between
 theoretical predictions for that part of the pion optical potential
 and the strength demanded by fits to pionic atoms data \cite{holinde}.
   The idea behind Ref.~\cite{wolfram} is that the isovector $\pi N$
  scattering length, $b_1$, involving the factor $f_\pi^{-2}$, will be
  enhanced in the nucleus\footnote{Similar effect will also occur for
  the isoscalar $\pi N$ scattering length, $b_0$, but in
  Ref.~\cite{wolfram} such an effect is neglected because of the very
  small value of $b_0$ deduced from pionic hydrogen and
  deuterium~\cite{Sc01}.}, renormalizing both the isovector part of
  the $s-$wave pion selfenergy, proportional to $b_1$, as well as the
  isoscalar part from the Pauli blocking rescattering correction,
  \cite{ericson}, which is proportional to ${b_0^2+ 2 b_1^2}$.
  
   A more detailed study was done recently in Ref.~\cite{friedman}
 where conducting fits to the pionic atoms data and re-scaling the
 $b_1$ free parameter by the ratio $(f_\pi/f_\pi(\rho))^2$, an
 improved agreement with the data was obtained. The agreement became
 excellent, without invoking any other extra repulsion, when the
 relativistic corrections of Ref.~\cite{birbrair} were also
 included. However, the relativistic corrections of Ref.~\cite{birbrair} were
 found to be ambiguous in Ref.~\cite{gal} and another study
 \cite{koch} showed that they were a consequence of approximations
 which broke the exact cancellation of some large terms, and no
 correction was found when the exact calculation was done. Leaving
 apart these relativistic corrections, the fact still would remain
 that the renormalization of $b_1$ would generate a considerable part
 of the missing repulsion, leading the author of Ref.~\cite{friedman} to
 claim that the data of pionic atoms offered an evidence of partial
 chiral restoration in nuclei.
 
   The problem is more subtle than just replacing the value of
   $b_1$. Actually, changing the $s-$wave real part of the optical
   potential is justified if one might argue that the replacement of 
$f_\pi$ by $f_\pi(\rho)$ is done at the level of the in--medium Chiral 
Lagrangian. But, then the changes in the
   $f_\pi$ parameter should be implemented in a theoretical
   calculation wherever this parameter appears.  This reminds us that
   this parameter actually appears not only in the
   $b_0\rho,~b_1\delta\rho$ and the Pauli blocking rescattering
   terms mentioned in Ref.~\cite{wolfram}, but in all terms of the optical
   potential, including the $p-$wave part of the potential, and more
   importantly, all the absorption 
terms\footnote{
   There is also another point worth mentioning since in the pion absorption
   terms \cite{juan,carmen,carmen1} the $\pi N$ amplitude appears half
    off-shell, with the off-shell pion with an energy of $m_\pi /2$  and a 
    momentum of 
   $\sqrt{m_\pi M_N}=360$ MeV approximately. Although we do not expect drastic
   changes with this moderate off-shellness, the renormalization of $b_1$ in
   this amplitude could be different than that for on--shell pions \cite{comment}. 
   However, this problem also appears in the Pauli corrected rescattering term
   mentioned above, which provides the main source of $s-$wave repulsion when
   the renormalization of $b_1$ is done. Indeed, the derivation of this term in
   a many body framework was done in \cite{carmen}, Fig.~22 and Eq.~(A7), and
   it was shown to come from rescattering terms implicit in a Lippmann Schwinger
   equation when the Pauli blocking in the intermediate nucleon states was
   considered. In this rescattering term one can see (Eq.~(A7) of
   \cite{carmen}) that the term also involves the half off-shell $\pi N$
   amplitude with the off-shell pion with a four-momentum $(q^0,q)$ such that 
   $m_\pi -E_F < q^0 <  m_\pi +E_F$ and $0<q<2k_F$, where $E_F$ and $k_F$ are
   the Fermi energy and momentum respectively. One can thus see that the level
   of off-shellness is similar in the rescattering term and in the absorption
   terms.  
}, which go like
   $f_\pi^{-6}$. 
We will show that this scenario is strongly disfavoured by the
   pionic atom data. These findings are corroborated by the recent
   works of Ref.~\cite{waki}, where a new mode of chiral
restoration is suggested in which the longitudinal $\rho$ would be the
chiral partner of the pion and both would become massless in the limit
of chiral symmetry restoration. These works show that the parameter
$f_\pi$ appearing in Eq.~(\ref{eq:frho}) and the one that appears in the chiral
Lagrangians when performing perturbation theory calculations with them,
let us call it $\hat{f}_\pi$, are not the same object. They would be
the same at tree level, but, as soon as perturbation theory is performed,
$\hat{f}_\pi$ becomes explicitly dependent on the scale of 
renormalization and is not equal to zero even in the limit of chiral
restoration where $f_\pi$ appearing in Eq.~(\ref{eq:frho}) would vanish. This line
of thought about the inadequacy of using the constant $f_\pi$ that one
induces from the GOR relation as the coupling constant in  a
perturbative approach, is in accordance with our ideas expressed below,
where we give different arguments on why the identification of these two
objects in a many body microscopic calculation would lead to
doublecounting. 
    
    In the present work we shall explore the consequences of
    replacing $f_\pi$ by $f_\pi (\rho)$ in all the terms where it appears in a
    theoretical evaluation of the pion selfenergy
\footnote{
At this point, it is of interest to remind that the derivation of the
renormalization of $f_\pi$ is linked
to the renormalization of the axial vector current. The links to the
renormalization of $f_\pi$ appearing in the $s-$ and $p-$waves of the
$\pi N$ scattering amplitudes are not so clear from the way the
derivation of $f_\pi (\rho)$ is done, and does not have to be the same
for $s-$ or $p-$waves \cite{comment}. 
}. 
   For the purpose of
    completeness we shall also investigate what happens in the case where the
    modifications are done in the $s-$wave but not in the $p-$wave parts. 
   However, while the idea of changing $f_\pi$ by its medium
   value, everywhere that it appears, seems reasonable with the caveats pointed
   above, it is also true
   that one has to look in detail in the elements of the theoretical
   derivation of the potential to avoid double-counting in the case
   that the many body approach already contains the renormalization
   mechanisms that would lead to the quenching of the $f_\pi$
   parameter\footnote{ 
It is worth mentioning that studies of the
   renormalization of the axial current in nuclei, using standard many body
   theory with effective Lagrangians, have been done both for the space
   component \cite{rho,towner} as well as for the time component
   \cite{kubodera,mariana}.
}.  
   
    The point we would like to make here is that while in the chiral
   studies the interpretation of the renormalization of the axial vector current
   as a change of the pion decay constant in the medium is a valid option, the
   standard many body approach in which the currents are renormalized using
   effective Lagrangians and the pions are renormalized in the medium using the
   same Lagrangians is also a valid option, but then one cannot reinterpret the
   renormalization of the axial vector current in terms of a change of the pion
   decay constant, recast the terms obtained for the pion selfenergy in terms of
   this decay constant and change it in all terms where it appears. A consistent
   many body approach using effective Lagrangians, as done for instance in
   \cite{juan,carmen,carmen1,osetweise,kaiserweise} would be a valid approach by itself.

   In this sense it is also interesting to mention that in the recent study 
   of Ref.~\cite{jose} the 
   corrections to the space part of the pion decay constant
    were  related to
   $\Delta $ excitations induced from the axial current, thus
   connecting with the findings of Refs.~\cite{rho,towner} where a
   conventional many body expansion, including nucleon and isobar
   degrees of freedom, was done. These are also the degrees of freedom used in 
   \cite{carmen,juan} to evaluate the pion selfenergy.
  
     All this said, the purpose of the present paper is to
  investigate, in the line of Ref.~\cite{friedman}, what would happen
  if we ignore the points discussed above and simply change $f_\pi$ by
  $f_\pi (\rho)$ in all the terms of the microscopic many body
  calculation of Ref.~\cite{juan}. The exercise is illustrative
  because although there might seem that pionic atoms are just
  governed mostly by the $s-$wave optical potential, this is not
  actually the case and the $p-$wave potential, as well as the
  absorption terms, responsible for the width of the pionic states,
  play also a very important role \cite{batty}.
  
\section{Results}

We analyze in this section, the partial chiral restoration effect on
the theoretical description of the pionic atom data. We base this
study on our previous and detailed work of Ref.~\cite{juan} and use
the same set of experimental pionic shifts and widths as in this
reference. There is a total of 61 piece of data. We present in
Table~\ref{tab:uno} the values of $\chi^2/N$, where $N=61$ is the 
number of data,
obtained with two different optical potentials ({\bf II} and
{\bf III}) modified to somehow account for the change of the pion
decay constant in the medium, together with the value obtained with
the theoretical potential ({\bf I}) developed in Ref.~\cite{juan}.

More specifically, potential {\bf I} in Table~\ref{tab:uno}
corresponds to the potential {\it TH} of Ref.~\cite{juan}, defined in
Eqs. (20-30) and (34-36) of that reference for the $p-$ and $s-$wave
parts of the optical potential. For the $s-$wave term the {\it TH} potential
uses
\begin{equation}
b_0 = -0.013 m_\pi^{-1} \qquad b_1 = -0.092 m_\pi^{-1} \qquad {\rm Im}
B_0 = 0.041 m_\pi^{-4} \label{eq:1993}
\end{equation}
where $b_0$ and $b_1$ were taken from the experimental analysis of
Ref.~\cite{Ho79} and ${\rm Im} B_0$ computed in
Ref.~\cite{carmen}.  The potential ({\it TH}) has been
developed microscopically and it contains the ordinary lowest order
optical potential pieces constructed from the $s$-- and $p$--wave $\pi
N$ amplitudes. In addition second order terms in both $s$-- and
$p$--waves, responsible for pion absorption, are also considered.
Standard corrections, as second-order Pauli re-scattering term, ATT
term, Lorentz--Lorenz effect and long and short range nuclear
correlations, are also taken into account. This theoretical potential
reproduces fairly well the data of pionic atoms (binding energies and
strong absorption widths)~\cite{juan} and low energy $\pi$-nucleus
scattering~\cite{juan2}. 

In the potential {\bf II} we take for the $p-$wave part that of
potential {\bf I}, but the $s-$wave part is modified following the
prescription of Ref.~\cite{wolfram}. Thus, we replace $b_0, b_1$ and ${\rm
Im} B_0$ by $F_\chi \times b_0, F_\chi \times b_1$ and $F_\chi^3
\times {\rm Im} B_0$ respectively, where $F_\chi$ can be deduced from
Eqs.~(\ref{eq:qqrho}) and ~(\ref{eq:frho}),
\begin{equation}
F_\chi  = F_\chi(\rho) = \left ( \frac{f_\pi}{f_\pi(\rho)} \right )^2 \approx 
\frac{1}{1-\sigma_N \rho/(f_\pi^2 m_\pi^2)} = \frac{1}{1-2.3 {\rm fm}^3 \rho} 
\end{equation}
for $\sigma_N = 50$ MeV and neglecting the in-medium pion mass change
because of its Goldstone boson nature. Note that 
$F_\chi(\rho)$ depends on the spatial coordinate $\vec{r}$ in the 
local density approach which we use for the optical potential.  
In potential
{\bf III} we modified both $s-$ and $p-$wave parts of the pion-nucleus
optical potential. Thus, in addition to the changes mentioned above
for the $s-$wave, in the computation of the $p-$wave part we have
replaced the $\pi N N$ and $\pi N \Delta$ coupling constants, $f$ and
$f^*$ in the notation of Ref.~\cite{juan}, by $\sqrt{F_\chi} \times f$
and $\sqrt{F_\chi} \times f^*$ respectively\footnote{For the case of
the in medium $\Delta-$selfenergy, $\Sigma_\Delta$, for which no 
explicit expression in terms of the $f$ and $f^*$ coupling constants
is given in Ref.~\cite{juan}, we have corrected only the leading
$\rho$ behaviour, this is to say, we have multiplied $\Sigma_\Delta$ by
$F_\chi^2$. Thus, for the imaginary part, we have not modified the
saturation coefficient appearing in the argument of the arc-tan
parametrization of Eq.~(11) of \cite{juan}.}. Note that, because of the
non-local nature of the $p-$wave part, first and second order
derivatives of $F_\chi(\rho)$ are needed. Neglecting those derivatives lead
to negative widths of the pionic atoms.
\begin{table}
\begin{center}
\begin{tabular}{c|c|c|c}\hline\tstrut\tstrut
 Potential & $\chi^2/N$ & $\chi^2_\epsilon/N_\epsilon$ & 
$\chi^2_\Gamma/N_\Gamma$ \tstrut\tstrut\tstrut\\\hline\tstrut\tstrut\tstrut
{\bf I}                             & 5  &   5     & 5 \\
{\bf II}                            & 47 &   64    & 30 \\
{\bf III}                           & 67 &   16    & 112\\\hline
\end{tabular}
\end{center}
\caption{Second column: values for $\chi^2/{\rm (num. data=61)}$
obtained for the potentials {\bf I, II} and {\bf III} defined in the text,
when the set of experimental pionic shifts and widths, used in
Ref.~\protect\cite{juan} are considered. Third and fourth columns:
Shift ($\chi^2_\epsilon = \sum_{\rm \epsilon}\left [(\epsilon^{\rm
pot}-\epsilon^{\rm exp})/\delta \epsilon^{\rm exp}\right]^2$) and
width ($\chi^2_\Gamma = \sum_{\rm \Gamma}\left [(\Gamma^{\rm
pot}-\Gamma^{\rm exp})/\delta \Gamma^{\rm exp}\right]^2$)
contributions to $\chi^2$ ($= \chi^2_\epsilon+\chi^2_\Gamma$), divided by
the number of shifts ($N_\epsilon=29$) and widths ($N_\Gamma=32$). 
The $s-$wave part of the optical potentials uses the parameters
given in Eq.~(\protect\ref{eq:1993}).}
\label{tab:uno}
\end{table}

The results of Table~\ref{tab:uno} clearly contradict the expectations of
Ref.~\cite{wolfram}, corroborated in Ref.~\cite{friedman}, of
finding a signal of partial chiral restoration in nuclei from pionic
atom data. 

To better understand the results presented in Table~\ref{tab:uno},
one should bear in mind that the real part of the $s-$ and $p-$wave
contributions to the pion-nucleon optical potential are repulsive and
attractive, respectively, and the fact that $F_\chi$ is a number
greater than one for all densities, i.e., the effect of the partial chiral 
restoration is
to increase, in absolute value, the size of each part ($s-$ and
$p-$waves) of the potential.  Thus, the potential {\bf II} is more
repulsive than the potential {\bf I}. The main source of the
enhancement of the repulsion in the $s-$wave 
part of the potential is through the increase of the Pauli blocking 
rescattering term, which is of isoscalar nature and goes, as quoted above,
as $b_0^2+2b_1^2$. Besides, the imaginary part of the 
potential {\bf II} is substantially bigger in absolute value than that of 
the potential {\bf I}, since it gets
multiplied by a factor $F_\chi^3$. Thus, we see in the table that the shifts of potential {\bf II}  are
clearly worse than those obtained with the potential {\bf I}. The same
occurs for the widths, those of  potential {\bf II} are bigger than the
experimental ones, but the effect is not as drastic as one might
expect since the potential {\bf II} is more repulsive than the potential
{\bf I} and thus the effective densities seen by the pion for the
potential {\bf II} case are smaller than those relevant when the
interaction {\bf I} is used. When the partial chiral restoration
effects are also incorporated to the $p-$wave, potential {\bf III},
the shifts are improved respect to the potential {\bf II} case. This is
because there is a cancellation between the increase of repulsion and
attraction generated by the inclusion of the chiral effects in each
wave of the optical potential. However, the potential {\bf III} has an
imaginary part too big, a fact which is clearly appreciated in the
table ($\chi^2_\Gamma/N_\Gamma = 112$).

However, a word of caution must be said now. The recent determination
of the isovector and isoscalar $\pi N$ scattering lengths, $b_1$ and
$b_0$, in Ref.~\cite{Sc01} 
\begin{equation}
b_0 = -0.0001^{+0.0009}_{-0.0021} ~m_\pi^{-1} \qquad 
b_1 = -0.0885 ^{+0.0010}_{-0.0021} ~m_\pi^{-1} \label{eq:2001}
\end{equation}
is incompatible, specially for $b_0$, with that of
Ref.~\cite{Ho79} quoted in Eq.~(\ref{eq:1993}). When one uses the central values of $b_1$ and $b_0$
given above, to re-compute ${\rm Im}B_0$, following the lines of
Ref.~\cite{carmen}, one gets
\begin{equation}
{\rm Im}B_0 = 0.0345 m_\pi^{-4} \label{eq:2001b}
\end{equation} 
These new values ( Eqs.~(\ref{eq:2001}) and~(\ref{eq:2001b})) for the
$s-$wave part of the optical potential lead to the results presented
in Tables~\ref{tab:dos}~and~\ref{tab:tres}. 
Neither potential {\bf I} nor potentials {\bf
II} and {\bf III} provide an acceptable description of the data. For the
interaction {\bf I} case, the main effect is the important reduction
of the $s-$wave repulsion which leads to a poorer description of
binding energies and to greater effective densities felt by the pion
than in the case presented in Table~\ref{tab:uno}. This latter effect
increases, in absolute value, the imaginary part of the optical
potential. On the other hand the smaller value of the parameter ${\rm
Im}B_0$ in Eq.~(\ref{eq:2001b}) than in Eq.~(\ref{eq:1993}) reduces,
in absolute value, the imaginary part. The total effect depends on the
pionic level ($nl$ and nucleus) considered. Partial chiral restoration
effects incorporated in potential {\bf II} increase the repulsion and
lead to an acceptable description of the shifts, but the increase in
the imaginary part, in absolute value, produces an unacceptable description
of the widths. The results presented in Ref.~\cite{friedman} correspond to
a situation like that of the potential {\bf II} in
Table~\ref{tab:dos} but where the imaginary part of the $s-$wave optical
potential is not being affected by this partial chiral
restoration. While this could be an acceptable procedure from the 
fitting point of view,  is not 
satisfactory from a theoretical or microscopical point of view.
  In any case, we have simulated this scenery by the
potential {\bf II$^*$}, which only scale by the chiral factor $F_\chi$ the
dispersive real part of the $s$--wave potential. This potential {\bf II$^*$}  is
the one providing the best description without any fitted parameter,
as can be seen in Table~\ref{tab:dos}, with also a large value of $\chi^2/N$=12. 
Finally, potential {\bf III} provides really poor results:
imaginary parts, in absolute values, are too big while the potential
turns out to be too attractive, once the partial chiral restoration
enhancement of the $p-$wave potential is also considered.
We consider adding a phenomenological fitted part to the previous
potentials {\bf I} and {\bf III} as done in \cite{juan}.
The results obtained are shown  in rows 
{\bf I$^{\rm fit}$} and {\bf III$^{\rm fit}$} of Table~\ref{tab:tres}.
 For potential 
{\bf I$^{\rm fit}$}, the fitted parameters are $\delta b_0=
 -0.0207(10)~m_\pi^{-1},\ \delta b_1= -0.0163(60)~m_\pi^{-1},\ \delta
 Im B_0=  0.0152(17)~m_\pi^{-4},\ \delta c_0=
0.045(10)~m_\pi^{-3},\ \delta c_1=  0.092(51)~m_\pi^{-3},\ \delta Im
C_0=  0.128(23)~m_\pi^{-6}$.
For potential 
{\bf III$^{\rm fit}$}, the fitted parameters are $\delta b_0=
 -0.0235(12)~m_\pi^{-1},\ \delta b_1= 0.0201(64)~m_\pi^{-1},\ \delta
 Im B_0=  -0.0455(24)~m_\pi^{-4},\ \delta c_0=
-0.078(14)~m_\pi^{-3},\ \delta c_1=  -0.040(56)~m_\pi^{-3},\ \delta Im
C_0= -0.212(36)~m_\pi^{-6}$.

   The results of the fits are instructive. In both cases we can see that there
 is a need for extra $s-$wave repulsion in the demanded value of $\delta b_0$,
 similar in both cases, and around a value 50 percent bigger than the repulsion
 provided by the old value of $b_0$, Eq.~(\ref{eq:1993}).  On the other hand, while from 
 the fit {\bf I$^{\rm fit}$} the data
 demand a value for $b_1$ about 16 percent bigger in size that the free value,
 the results of  {\bf III$^{\rm fit}$} are telling us that that the
 renormalization of $f_\pi$ provides an effective value of  $b_1$ about 
 20 percent larger than demanded by the data. The effects on $Im B_0$ are more
 striking. While the data would demand a value about 30 percent bigger than the
 one provided by the theoretical potential {\bf I}, the effectively renormalized
 value of $Im B_0$ in potential {\bf III} requires a reduction three times
 bigger. These results are thus telling us again that largely renormalized 
 values of $b_1$  are not welcome by the data. Particularly, the values of
 $ImB_0$ that are obtained after multiplying by $F_\chi^3$ the theoretical
 value of $ImB_0$ are far too large.
\begin{table}
\begin{center}
\begin{tabular}{c|c|c|c}\hline\tstrut\tstrut
 Potential & $\chi^2/N$ & $\chi^2_\epsilon/N_\epsilon$ & 
$\chi^2_\Gamma/N_\Gamma$ \tstrut\tstrut\tstrut\\\hline\tstrut\tstrut\tstrut
{\bf I}                             & 84  &  158   &  16 \\
{\bf II}                            & 47  &    4    & 86  \\
{\bf II$^*$}                            & 12  &    16    & 8  \\
{\bf III}                           & 207 &  200    & 215 \\\hline
\end{tabular}
\end{center}
\caption{Same as in Table~\protect\ref{tab:uno}, but for the $s-$wave part of the
optical potentials, the parameters given in
Eqs.~(\protect\ref{eq:2001}) and~(\protect\ref{eq:2001b}) have been
used. The potential II$^*$ is like the potential II (the 
The $s$--wave coefficients $b_0$ and
$b_1$ are replaced by $F_\chi \times b_0$ and  $F_\chi \times b_1$),
except that  the $s$--wave absorptive part, Im$B_0$, is kept unchanged
as in the potential I.}
\label{tab:dos}
\end{table}
\begin{table}
\begin{center}
\begin{tabular}{c|c|c|c}\hline\tstrut\tstrut
 Potential & $\chi^2/N$ & $\chi^2_\epsilon/N_\epsilon$ & 
$\chi^2_\Gamma/N_\Gamma$ \tstrut\tstrut\tstrut\\\hline\tstrut\tstrut\tstrut
{\bf I$^{\rm fit}$}                  & 1.8  &  0.8   & 2.8 \\
{\bf III$^{\rm fit}$}               & 3.1 &  1.6    & 4.4 \\\hline
\end{tabular}
\end{center}
\caption{Results from potentials {\bf I$^{\rm fit}$} and {\bf III$^{\rm fit}$}
obtained from a best fit to the data after adding  phenomenological terms to
potentials {\bf I}  and {\bf III} of table 2.}
\label{tab:tres}
\end{table}

\section{Conclusions}

  The idea expressed here is that, without questioning the dropping of
  the pion decay constant in a nuclear medium, a systematic many body
  expansion using effective Lagrangians, consistent with present
  knowledge of $\chi PT$ and incorporating explicitly the elements
  which go into the counter-terms of $\chi PT$, ( for instance 
  $\Delta h$ excitations), already incorporates the
  mechanisms responsible for the dropping of $f_\pi$ in the medium and,
  hence, this explicit renormalization should not be considered in
  addition to avoid double-counting.
  
    It should also be stated that efforts were made to solve the
  puzzle of the missing repulsion and several small corrections were
  found to the $s-$wave selfenergy, among them corrections of second
  order in the density from dispersion corrections linked to the
  $s-$wave pion absorption \cite{carmen}, from the pion scattering with
  the virtual pion cloud in the nucleus 
  \cite{rockmore, carmen2,kaiserweise,park} and
  from a Lorentz- Lorenz correction to the $s-$wave rescattering terms 
  \cite{holinde}, such that taking into account all the different
  effects and their errors there was only moderate room for extra
  $s-$wave repulsion \cite{holinde}. Of course, the new data for the 
  $b_0$ parameter have reduced a source of $s-$wave repulsion provided 
  by the old data of \cite{Ho79} and have worsened the problem, such 
  that we can say that with the values of the scattering lengths provided 
  by the recent data of pionic hydrogen \cite{Sc01} the problem of the missing 
  repulsion in pionic atoms would be reopened.  
  
    The fits conducted, before and after the renormalization of $f_\pi$ is done,
  are also telling us that the strong renormalization of $b_1$ provided by 
  this idea is not welcome by the data. Particularly the value for $ImB_0$ which
  is multiplied by $F_\chi^3$. 
  
  All together the conclusions of this paper seem clear: First, that
  more theoretical work needs to be done to understand the meaning of
  the dropping of the pion decay constant and how it could have a
  repercussion in the renormalization of the $\pi N$, $\pi NN$, etc.,
  amplitudes in the nuclear medium within a determined many body
  scheme. And second, that it is quite important to settle the
  question of the precise values of the $\pi N$ scattering lengths, so
  that the present discrepancy between the results from pionic
  hydrogen and from scattering data, reminiscent of the one that
  remained for long in the $K^- p$ problem, and which was finally
  settled in \cite{iwasaki}, should be resolved. In this respect it is
  interesting to mention that efforts in this direction are presently
  underway. As an example the recent paper \cite{deloff}, which makes
  a combined analysis of pionic hydrogen and deuterium data, advocates
  for values of the $b_0$ parameter of the order of $-$0.004
  $m_\pi^{-1}$, which are much bigger in strength than the value
  quoted here in Eq.~(\ref{eq:2001}). Should however results of the
  order of those used in Eq.~(\ref{eq:2001}) prevail, then it is clear
  that there is an important missing $s-$wave repulsion which again
  will require renewed theoretical efforts to be understood.

\subsection*{Acknowledgments}

 We would like to acknowledge some useful
discussions with J.A. Oller and useful comments from W. Weise and E. Friedman.
 This work is partly supported by DGICYT contract number BFM2000-1326, by
DGES contract number PB98-1367, by the Junta de Andalucia and 
by the EU TMR network Eurodaphne, contract no. ERBFMRX-CT98-0169.

\end{document}